\documentclass[pra,twocolumn,aps,showpacs,amsmath,amssymb]{revtex4-1}

\usepackage{color}
\usepackage{bm}
\usepackage{graphicx}
\usepackage{braket}

\usepackage[plainpages=false,pdfpagelabels,colorlinks=true,linkcolor=red,urlcolor=blue,citecolor=blue,pdftitle={Title},pdfauthor={},pdfdisplaydoctitle=true,pdfduplex=DuplexFlipLongEdge]{hyperref}

\def\Tr{\mbox{Tr}\,}

\newcommand\ba{\begin{eqnarray}}
\newcommand\ea{\end{eqnarray}}
\newcommand\be{\begin{equation}}
\newcommand\ee{\end{equation}}

\begin{document}

\title{Many-body dynamical phase transition in quasi-periodic potential}
\author{Ranjan Modak$^1$ and Debraj Rakshit$^1$} 
\affiliation{$^1$ Department of Physics,Institute of Science, BHU, Varanasi-221005, India}
\begin{abstract}
Much has been learned regarding dynamical quantum phase transition (DQPT) due to sudden quenches across quantum critical points in traditional quantum systems. However, not much has been explored when a system undergoes a localization-delocalization transition. Here, we study one dimensional fermionic systems in presence of a quasi-periodic potential, which induces delocalization-localization transition even in 1D. We show signatures of DQPT in the many-body dynamics, when quenching is performed between phases belonging to different universality classes. We investigate how the non-analyticity in the dynamical free energy gets affected with filling fractions in the bare system and, further, study the fate of DQPT under interaction. Strikingly, whenever quenching is performed from the low-entangled localized phase to the high-entangled delocalized phase, our studies suggest an intimate relationship between DQPT and the \emph{rate} of the entanglement growth -- Faster growths of entanglement entropy ensures quicker manifestation of the non-analiticties in the many-body dynamical free energy.

\end{abstract}
\maketitle

\section{Introduction}
Isolated out-of-equilibrium systems posses many challenges, and concurrently, offers new possibilities: Whereas it gives rise to complex physical phenomena beyond the reach of equilibrium statistical mechanics, it also opens door for discerning equilibrium statistical properties in an unique manner \cite{Eisert15}. Dynamical quantum phase transition (DQPT) \cite{Heyl18} constitutes a prime example that has illuminated a new light on the traditional understanding of quantum phase transition, which is developed on the pillars of equilibrium statistical mechanics.

DQPT is built upon by drawing parallel ideas borrowed from equilibrium physics, where Loschmidt amplitude, $\mathcal{L}(t)$, does a similar kind of business that the partition function does in the equilibrium physics of quantum phase transition. $\mathcal{L}(t)$ quantifies the  overlap between an initial state and the time-evolved state, and is defined as $\mathcal{L}(t) = \langle \psi_i| \exp(-i \hat{H} t)|\psi_i \rangle$, where $|\psi_i \rangle$ is an initial state, and $\hat{H}$ is the driving Hamiltonian. The signature of a dynamical phase transition is imprinted in form of the non-analyticities at certain time instances, $t=t^*$, in the dynamical free energy, $f(t)=-2 \lim_{L \to \infty} \ln |\mathcal{L}(t)|/L$.

DQPT has been investigated elaborately in various contexts of quantum phase transitions in quantum model systems \cite{Heyl13, Sedlmayr19, Karrasch13, Andraschko14,Halimeh18,Mishra18,Shpielberg18,Srivastav19,Zunkovic18,Hagymasi19,Huang19,Soriente19,Cao19,Gurarie19,Abdi19,Haldar20}, and also for topological transitions \cite{Vajna15,Schmitt15,Jafari17, Sedlmayr18,Jafari18,Zache19}. 
Along with sudden quench, periodically driven systems have been studied as well \cite{Sharma14,Yang19}. Although DQPT is mostly studied with the pure ground state as an initial state,  conceptual generalization has been established in the cases of degenerate states, mixed state, or even the open systems \cite{Abeling16, Mera17,Bhattacharya17,Heyl17,Lang18a, Lang18b,Kyaw18}. Importantly, recent experimental advances with quantum simulators \cite{Lewenstein07,Lewenstein12,Georgescu14}, that can mimic real-time dynamics of isolated quantum many-body systems, justifies such theoretical investigations. DQPT has already been experimentally verified in laboratory \emph{via} ultracold atom \cite{Smale19}, and ion traps \cite{Jurcevic17}.

So far, conventional quantum phase transitions governed by Landau theory of spontaneous symmetry breaking have enjoyed the bulk amount of attention. This work, instead, considers localization-delocalization transition, which cannot be described by simple Landau theory. Localization-delocalization transition has been a subject of intense research since the seminal work of Anderson (1958), which revealed that despite the quantum tunneling processes a quantum particle may get localized in presence of a disorder \cite{Anderson58, Abrahams79}. Many-body localization (MBL) further considers the effect of interaction in addition to disorder, and off-late, has become a hot research topic \cite{Abani17,Alet18,Abanin19,Nandkishore6}. Localization-delocalization transition separates the localized phase from the thermal one, and a question that naturally arises is whether the equilibrium physics of localization-delocalization transition leaves a trace on the real-time dynamics, and if  the progressing time would play a role similar to the equilibrium control parameter, i.e. the strength of randomness.

Several of the early works with Anderson localization and MBL were carried on disordered quantum many-body systems, such as spin chains in presence of random field or fermionic lattice systems with random onsite potential. Another closely related setting considers quasi-periodic many-body systems \cite{Aubry80, Modugno09}, also known as the the Aubry-Andr{\'e} (AA) model, where  instead of pure randomness the incommensurate on-site potential  drives a system in the localized phase.   
In this model, the localization-delocalization transition occurs for a finite incommensurate potential amplitude, say $\Delta=\Delta_c$ \cite{Machida86,Geisel91,Roscilde08,Roux08,Deng08,Albert10,Cai10,He12,Gullo15}. This is different from the usual Anderson localization in one-dimension, which requires only an infinitesimal disorder strength to localize all states. It also has been shown that by introducing the interaction in such systems, the ergodic-MBL transition  takes place at a critical amplitude $\Delta > \Delta_c$ \cite{Iyer13}. Quasi-periodic on-site potentials can be engineered in ultracold simulators in a controlled manner \cite{Schreiber15,Luschen17,Luschen18}.

There has been a very recent effort in understanding the fate of DQPT in the context of the single-particle localization-delocalization transition in case of Aubrey-andre model \cite{Yang17},  where an  analytical expressions of the transition times ($t^*$) can be derived for a limiting situation of end-to-end quench between two distinct phases. In order to build an in-depth understandings of the \emph{many-body aspects} of the DQPT in localization-delocalization transition, we consider   fermions and probe the AA model at finite filling fractions. Owing to the added complexity associated with many-body generalization, we perform time extension of density matrix renormalization group (tDMRG) calculations \emph{via} matrix product state (MPS) formalism ~\cite{white-2004, daley-2004, uli-2011}. We investigate the appearances of the non analytic points and identify their trends for varied filling fractions, interaction strength and initial states. Similar to the single-particle calculation, we also find that the nonanalytic points in dynamical free energy  bearing the signatures of the DQPT appear only if an sudden quench is performed across the transition point, i.e. the system is suddenly driven from the localized (delocalized) to the delocalized (localized) phase.
Moreover, we investigate entanglement entropy, which often offers useful insights on quantum many-body phenomenology \cite{Eisert10,Amic10}. In particular, it has been recently suggested that the DQPTs may breed enhanced entropy production around the transition times \cite{Jurcevic17,Hagymasi19}. We find this to happen whenever the system is driven from a low-entanglement (localized) regime to high-entanglement (delocalized) regime. In-fact, our results provide solid evidence, which suggest that occurrences of DQPTs are intimately related to the \emph{rate} of information spreading, i.e. the entanglement growth rate. However, this may not be uni-vocally true, particularly, decisive statements cannot be passed for a sudden quench occurring other way around as highly-entangled initial states are driven by low-entangled Hamiltonians, hardly any entanglement production occurs.  It's now possible to measure entanglement entropy in cold-atom experiments \cite{Islam15,Kaufman16}, and hence our results can be experimentally verified.

Here is how the rest of the paper is organized. In Sec. II, we introduce the model. Section III discusses DQPT in fractionally filled non-interacting systems. In-particular, we concentrate on the situation when quenching is performed between phases belonging to different universality classes, i.e. the system is suddenly quenched from the (delocalized) localized to the (localized) delocalized phase. Moreover, we also study entanglement production associated with DQPT. In Sec.~IV, we investigate the effects of interactions. Finally, we draw conclusions in Sec. V.

\section{Model}
We study a system of fermions  in an one-dimensional lattice of size $L$, which is described by  the following  Hamiltonian:
\begin{eqnarray}
\hat{H}(\Delta)&=&-\sum_{i=1}^{L-1}(\hat{c}^{\dag}_i\hat{c}_{i+1}+\text{H.c.})+\Delta\sum_{i=1}^{L} \cos(2\pi\alpha i)\hat{n}_i  \nonumber \\
&+&V\sum_i\hat{n}_i\hat{n}_{i+1},
\label{nonint_model}
\end{eqnarray}
where $\hat{c}^{\dag}_i$  ( $\hat{c}_i$) is the fermionic creation (annihilation) operator at site $i$, $\hat{n}_i =\hat{c}^{\dag}_i\hat{c}_{i}$ is the number operator, and  $\alpha$ is an irrational number. Without loss of any generality, we choose $\alpha=\frac{\sqrt{5}-1}{2}$ for all the calculations presented in this work. In the absence of interaction i.e. $V=0$, the Hamiltonian $\hat{H}$ is known as Aubry-Andr{\'e} (AA) model. It  supports a  delocalization-localization transition as one tunes $\Delta$. In the thermodynamic limit, $\Delta=2$ corresponds to the transition point \cite{Aubry80}.

For the most calculations in this paper, the system initially prepared in a ground state $|\psi_i\rangle$ of the Hamiltonian $\hat{H}$ for $\Delta=\Delta_i$. We study subsequent unitary dynamics of the initial state following a sudden quench: $\hat{H}(\Delta_i) \to \hat{H}(\Delta_f)$. The time evolved state  is given by $|\psi(t)\rangle =e^{-i\hat{H}(\Delta_f)t}|\psi_i\rangle$. In order to detect DQPT, we focus on dynamical free energy, which is described by, 
\begin{equation}
    f(t)=-\lim_{L\to \infty}\frac{2}{L}\ln|\langle\psi(t)|\psi_i\rangle|.
\end{equation}
 It has been recently reported that DQPT occurs in single particle Aubry-Andr{\'e} model, once the quenching is performed from the localized phase to the delocalized phase or vice-versa ~\cite{Aubry80}.
Here, our interest is to understand the fate of the DQPT when finite number of particles are loaded in such systems. The filling fraction is identified as $\nu=N/L$, where $N$ is the total number of fermions. Given that we use tDMRG method for all our calculations,  we restrict our-selves to the open boundary condition.

\section{Fractionally filled bare systems}
Given that our main aim is to understand the \emph{many-body aspects} in DQPT for localization-delocalization transition, we first focus on the Hamiltonian $\hat{H}$ in the absence of interaction i.e. $V=0$. Subsequently, we study two distinct cases - quenching the Hamiltonian from (A) the delocalized to localized phase and (B) the localized to the delocalized phase.

\begin{figure}
\includegraphics[width=0.45\textwidth]{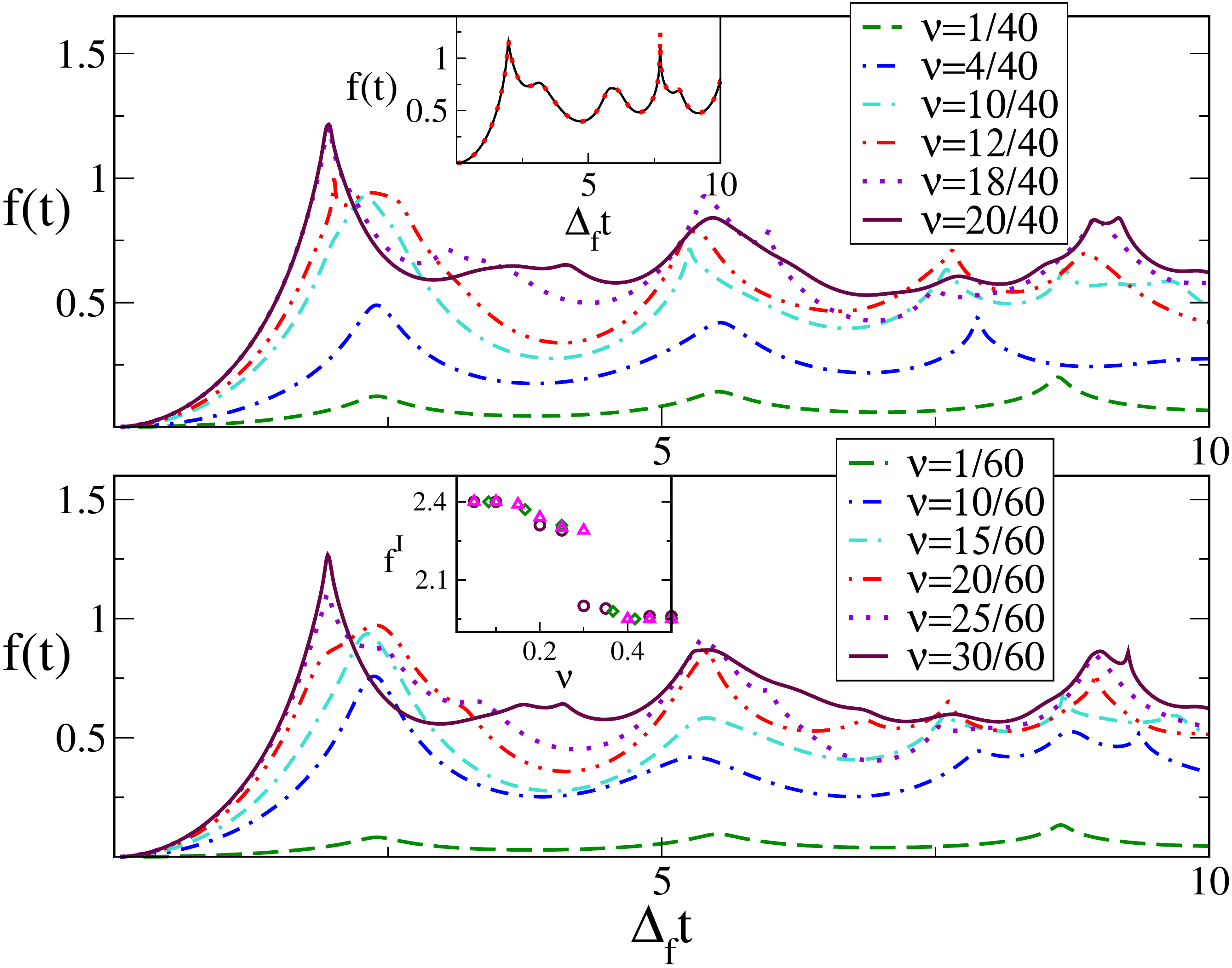}
\caption{\emph{Delocalized to localized quench -- Varied filling fractions.} Variation of the dynamical free energy $f(t)$ as a function of time $t$ for different values of the filling fractions, $\nu$, for (a) $L=40$ and (b) $L=60$. In the inset of (a) we bench mark our tDMRG results with exact diagonalization calculations for $L=16$ at half filling. Inset of (b) shows the variation of the $1^{st}$ non-analytic point $t^*$ with filling fraction $\nu$, where  the  triangle, circle, and diamond symbols correspond to $L=40$, 60, and 80 respectively. We choose $\Delta_i=0.2$ and $\Delta_f=100$.}
\label{fig1}
\end{figure}

\subsection{Delocalized to localized case}
 In this section we primarily investigate  the behavior of dynamical free energy $f(t)$ as we quench the Hamiltonian ~\eqref{nonint_model} from the delocalized phase to the deep into localized phase, i.e. $\Delta_i=0.2 \to \Delta_f=100$. Figure \ref{fig1} shows the variation of $f(t)$ with time for different values of the filling fractions. Two different system sizes are studied $L=40$ (Fig.~\ref{fig1}(a)) and $L=60$ (Fig.~\ref{fig1}(b)). In the single-particle limit, an analytical expression for $f(t)$ can be derived in the thermodynamic limit, $f(t)=-\frac{2}{L}\ln|J_0(\Delta_ft)|$, where $J_0$ is is the zero-order Bessel function,  for end-to-end quench with $\Delta_i=0  \to \Delta_f=\infty$ ~\cite{Yang17}. It implies that the non-analytic points in the dynamical free energy appear at $t^{*}_c=x_c/\Delta_f$, where $x_c$ with $c=1$, 2, 3 .. are zeroes of $J_0(x)$. Even for our choice of quench parameters (where $\Delta_i=0.2$ and $\Delta_f=100$), it turns out that the non-analytic points are $t^{*}_c\simeq x_c/\Delta_f$ for $N=1$. However, the non-analytic points in $f(t)$ start shifting with increasing number of particles. Given that in this work we primarily concentrate on  the 1st non-analytic point, it turns out that as we increase the filling fraction $\nu$, the 1st non-analytic point starts moving towards smaller values of $t^{*}$. We find that while for the single particle case, $\Delta_f t^* \simeq 2.41$, 
for $\nu=1/2$, $\Delta_ft^*\simeq 1.95$. In the inset of Fig.~\ref{fig1}(b), we show the positioning of the $1^{st}$ non-analytic point as a function of filling fraction for different system sizes, $L$ = 40, 60 and 80. 

\begin{figure}
\includegraphics[width=0.48\textwidth]{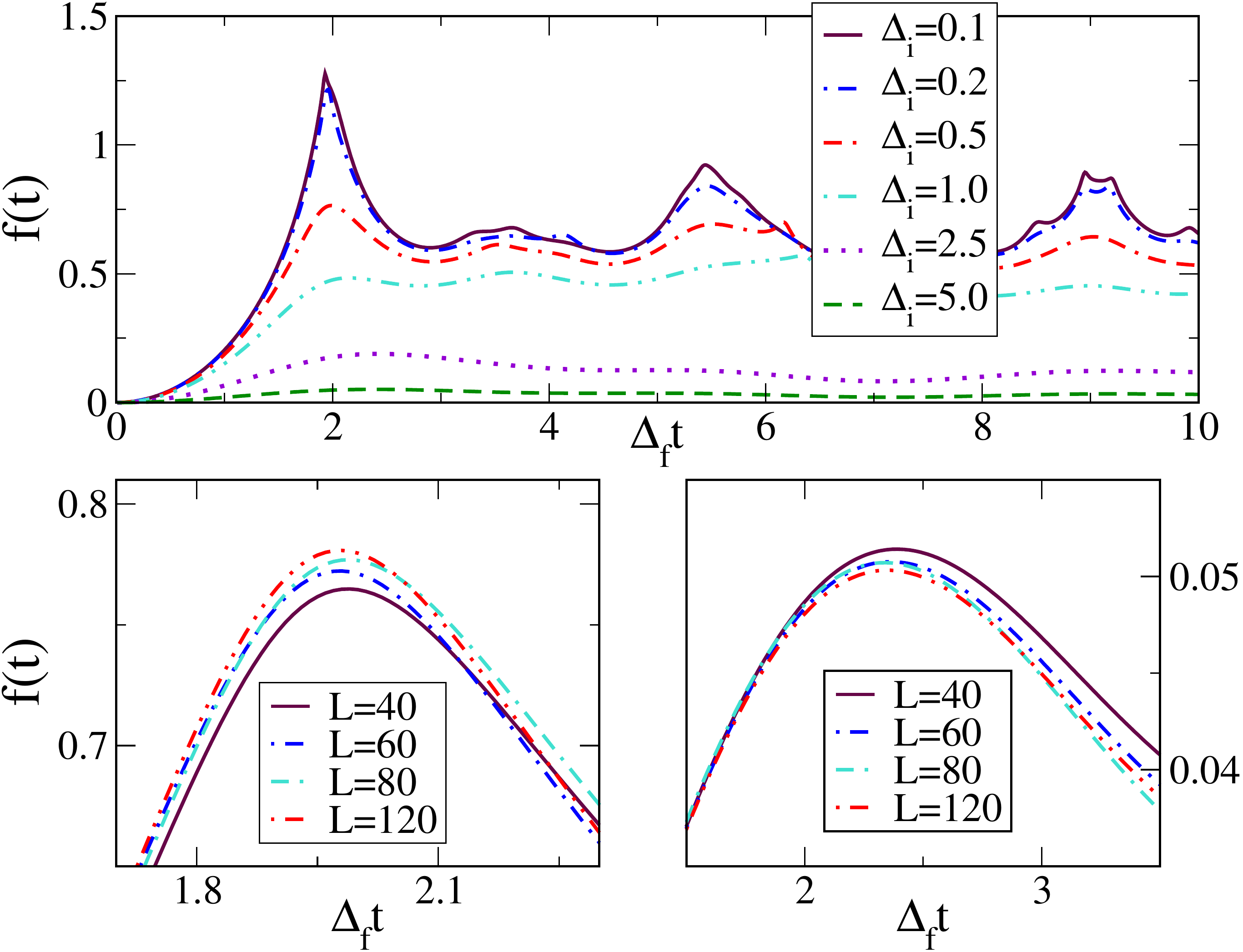}
\caption{\emph{Delocalized to localized quench -- Varied initial states.} (a) shows the variation of the dynamical free energy as a function of time for a fixed value of $\Delta_f$, but for different choices of $\Delta_i$. We choose $\Delta_f=100$, $L=40$, and $\nu=1/2$. Analysis of the finite size effects are presented in (b) for $\Delta_i=0.5$ and in (c) for $\Delta_i=5.0$, which monitor the change in the sharpness of the $1^{st}$ peak of the dynamical free energy with increasing system size. The chosen parameters are again $\Delta_f=100$, and $\nu=1/2$.}
\label{fig2}
\end{figure}

Next, we analyze finite size effects on the $1^{st}$ non-analytic point. For this, first we consider different values of $\Delta_i$, while keeping the driving Hamiltonian $(\Delta_f=100)$, filling fraction $(\nu=1/2)$, and the system size $(L=40)$ fixed. Fig.~\ref{fig2} (a) demonstrates variations of $f(t)$ with $\Delta_f t$ for different choices of the initial states, which are determined by setting $\Delta_i =$ 0.1, 0.2, 0.5, 1.0, 2.5, and 5.0. It is evident that the cases with $\Delta_i=0.1$ and $0.2$ are characterized by presence of "cusps" in time dynamics, providing clear signatures of the non-analyticities. However, the non-analytic features are not very clear for other values of $\Delta_i$. In contrast they appear like "humps".

In fig.~\ref{fig2} (b) and fig.~\ref{fig2} (c), we show the variation of these "humps" as we increase the system size $L$, for $\Delta_i=0.5$ and $\Delta_i=5.0$, respectively. We find that while for $\Delta_i=0 .5$ the "hump" is getting sharper as we increase $L$ (this feature has been observed for other values of $\Delta_i$ as well as long as $\Delta_i \lesssim 2$). On the other hand for $\Delta_i=5.0$, the peak flattens with increasing $L$. The trends clearly indicate that in the thermodynamic limit the non-analytic behaviour survives when the quench is performed  from the delocalized phase to localized phase. However, when  we quench the Hamiltonian ~\eqref{nonint_model} without crossing the phase boundary i.e. $\Delta_i>2$, the signature of non-analytic behaviour in $f(t)$ tends to wash away.  Similar conclusions also has been achieved for $N=1$ as well in Ref.~\cite{Yang17}.

\subsection{Localized to delocalized case}
\begin{figure}
\includegraphics[width=0.48\textwidth]{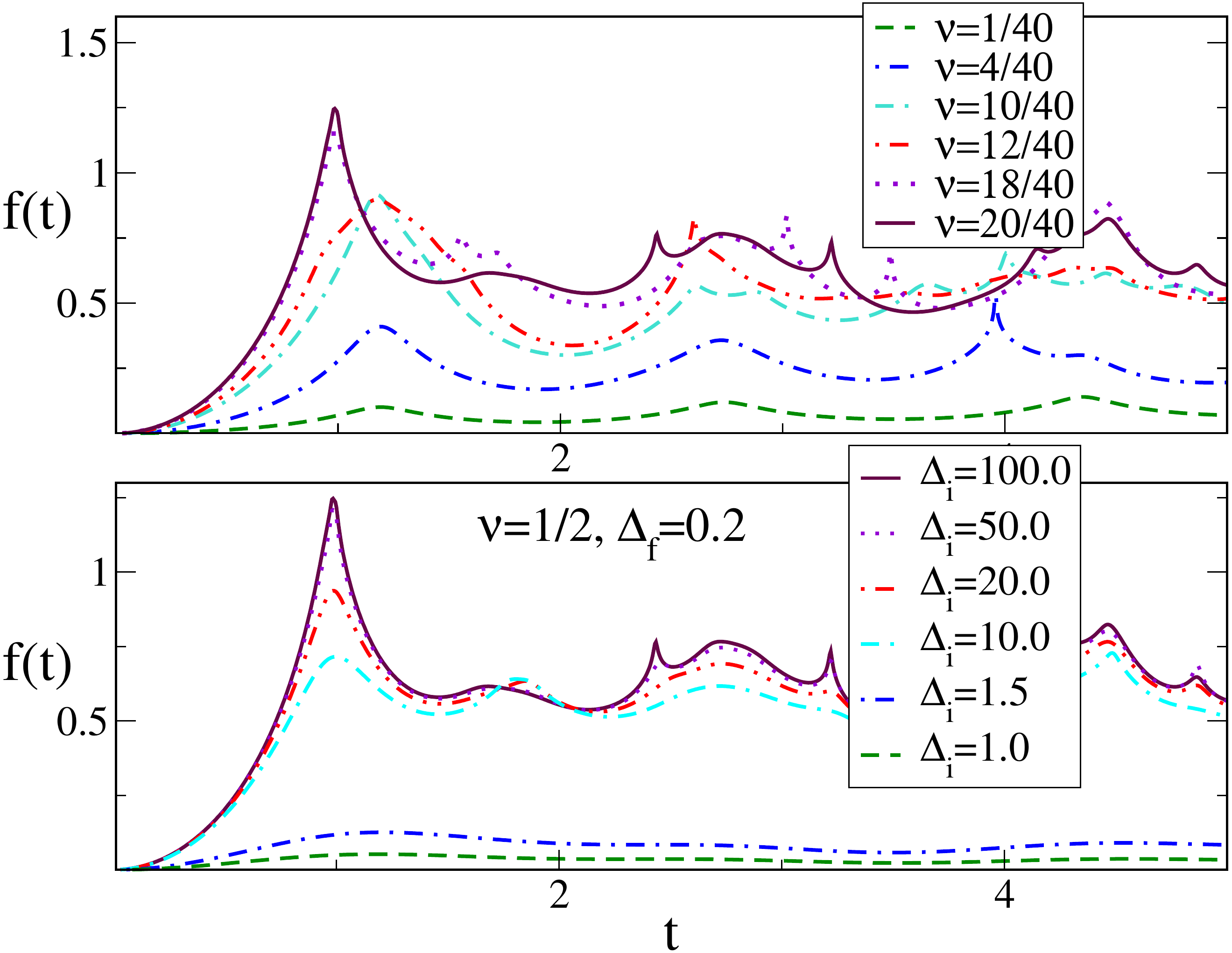}
\caption{\emph{Localized to delocalized quench.} Variation of the dynamical free energy $f(t)$  with time $t$ is shown in (a) for different filling fractions, $\nu$ and (b) for different initial states. Inset of (a) shows the variation 1st non-analytic point $t^*$ with $\nu$, where $\Delta_i=100$ and $\Delta_f=0.2$.  (b) shows the Variation of $f(t)$ with $t$ for $\nu=1/2$, $\Delta_f=0.2$ for different values of $\Delta_i$. $L=40$ for all the cases.}
\label{fig3}
\end{figure}

Now we focus on a quench from the localized phase to the delocalized phase.  Figure~\ref{fig3} shows the results for a quench from $\Delta_i=100$ 
$\to \Delta_f=0.2$. 
For the single-particle case i.e. $N=1$, it is straight froward to analytically track the non-analytic points in $f(t)$ corresponding to a quench  from $\Delta_i=\infty \to \Delta_f=0$. The non-analytic points in $f(t)$ corresponds to $t^*=x_c/2$, where $x_c$ are zeroes of $J_0(x)$ ~\cite{Yang17}.  Even for our choice of quench parameters i.e. $\Delta_i=100$ 
$\to \Delta_f=0.2$, our single-particle numerical results agrees well with the analytical results for the transition times, $t^{*}\simeq x_c/2$. However, as we increase the filling fraction,  $t^{*}$ starts relocating. This is shown in the inset of fig.~\ref{fig3} (a). We further show, by varying $\Delta_i$, while keeping $\Delta_f=0.2$ and filling fraction $\nu=1/2$ fixed, that if we quench from the deep into the localized phase to the delocalized phase, the non-analytic features become much sharper, at least for our choice of system size (see Fig.~\ref{fig3} (b)). On the other hand, when $\Delta_i \lesssim 2$, the signature of non-analyticity in $f(t)$ fades away. This is once again an evidence of  absence of the DQPT if the quench is performed within the same phase. 

\subsection{Entanglement production}
\begin{figure}
\includegraphics[width=0.48\textwidth,]{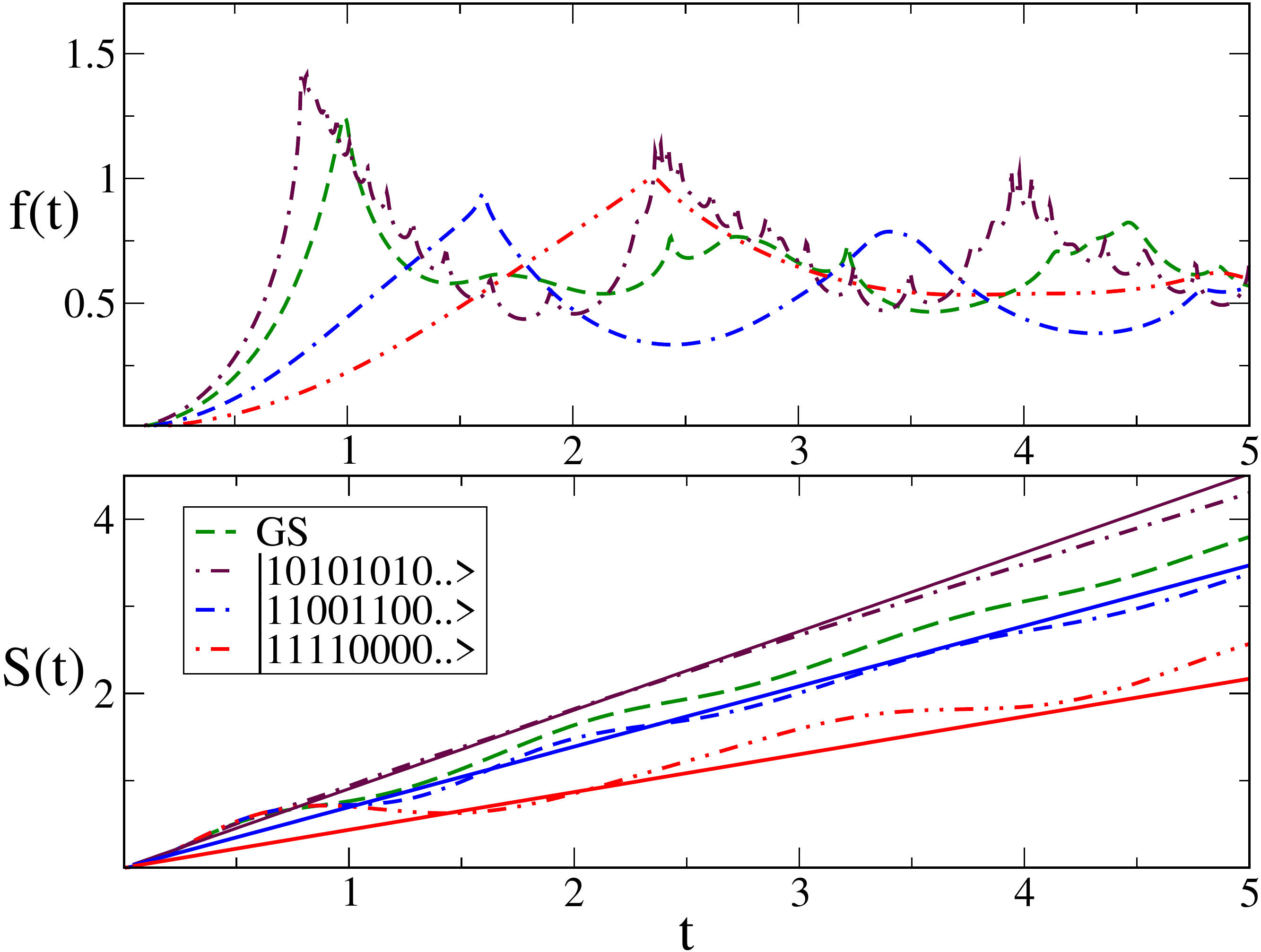}
\caption{\emph{Entanglement production and quasi-particle predictions.}(a) Dynamical free energy as a function of time for different initial product states for $L=40$ and $\Delta_f=0.2$ at half filling. 
(b) shows the entanglement entropy for the same set of initial states and parameters. Solid lines correspond to quasi-particle predictions.}
\label{fig4}
\end{figure}

In this section, we investigate the connection between the DQPT and the entanglement production. 
Given that here we quench from extremely deep localized phase ($\Delta_i=100$) to the delocalized phase, the initial state can be approximated quite accurately by a suitable product state. We have checked this explicitly by choosing two states, 1) the exact ground state of the pre-quench Hamiltonian 
$\hat{H}(\Delta_i=100)$, and 2) approximated product state obtained by filling desired number (depends on the total number of particles in the system) of lowest energy levels of the Hamiltonian  $\hat{H}_0=\sum_{i=1}^{L} \cos(2\pi\alpha i)\hat{n}_i$. We find that the results of $f(t)$ vs $t$ for these two states are almost indistinguishable.  
Hence, in Fig.~\ref{fig4} we don't restrict ourselves to only the ground state of the pre-quench Hamiltonian. We investigate the variation $f(t)$ with $t$ for different initial product state (e.g. Neel state). Given that  all these product states have extremely high overlap with some excited eigenstates of the pre-quench Hamiltonian $\hat{H}(\Delta_i=100)$, temporal variation of $f(t)$ for such states are expected to be very similar to some highly excited energy eigenstates of the Hamiltonian  $\hat{H}(\Delta_i=100)$.
Figure.~\ref{fig4}(a) shows that for many such states indeed one can find the clear signature of the DQPT. However, the location of the  non-analytic points $t^{*}$ varies quite a lot as we change our initial states. 

Next, we study the entanglement dynamics for the same set of initial states once we let it evolve under the unitary evolution governed by a
Hamiltonian belonging in the delocalized phase $\hat{H}(\Delta_f=0.2)$. Given that the dynamics is unitary, the entire system remains in a pure state. The reduced density matrix $\rho_A$ of a finite  subsystem A of length  $L/2$ is defined as $\rho_A = \Tr_B|\psi(t)\rangle \langle \psi(t)|$, where the trace is over the degrees of freedom of the complement B of A.  Here we only restrict ourselves to one of the most useful entanglement measures, the von Neumann (entanglement) entropy $S=\Tr[\rho_A\ln \rho_A]$.

Figure~\ref{fig4}(b) shows the results for the entanglement growth for the same set of initial states for which $f(t)$ has been plotted in the fig.~\ref{fig4} (a). We find an extraordinary correlation, i.e. the appearance of the non-analytic point (at least the 1st point) $t^{*}$ depends on the entanglement growth rate. At least 
among the choices of our initial state it seems that the entanglement growth rate for the Neel state i.e. $|\psi(0)\rangle=\Pi_{i=1}^{L/2}\hat{c}^{\dag}_{2i}|0\rangle=|1010..\rangle$ is maximum and where as, the entanglement growth rate of  $|\psi(0)\rangle=|11110000..\rangle$ is minimum. That is also reflected in the $f(t)$ vs $t$ plot in fig.~\ref{fig4} i.e.  the value of $t^{*}$ for the Neel state is much smaller compare to  other states. Intuitively we would expect that as well. Given that the non-analytic points in $f(t)$ corresponds to the zero fidelity $\mathcal{L}(t)$ in the thermodynamic limit, we would expect the speed at which the  memory of the initial state will be washed away should also depends on the entanglement growth rate. If the propagation rate of quantum correlation is higher, fidelity is expected to vanish in a much shorter time compared to the case where the rate of propagation of quantum correlation is smaller. This is precisely what has been manifested in Fig.~\ref{fig4}. 

Now given that the quench Hamiltonian $\hat{H}(\Delta_f)$ is integrable in the limit $\Delta_f\to 0$, the entanglement dynamics also can be described extremely efficiently within quasi-particle picture ~\cite{calabrese-2005}.  To understand why this is the case, let us first briefly describe the quasiparticle picture for the entanglement
spreading which is applicable to generic integrable models.
According to this picture, the initial state acts as a source of quasiparticle excitations which are produced in pairs
and uniformly in space. After being created, the quasiparticles move ballistically through the system with opposite
velocities. Only quasiparticles created at the same point in space are entangled, and while they move far apart, they carry forward entanglement and correlation in the system. A pair contributes to the entanglement entropy at time $t$ only if
one particle of the pair is in A (of size $\ell$) and its partner is in B. Keeping track of the linear trajectories of the particles, it is easy to conclude  ~\cite{calabrese-2005,alba-2016}
\begin{equation}
    S(t)=\sum_n 2t\int_{2|v_n|t \langle \ell}v_n(k)s_n(k) dk+\ell \int_{2|v_n|t \rangle \ell}s_n(k) dk.
    \label{qp}
\end{equation}

Here the sum is over the species of particles $n$ whose number depends on the model, $k$ represents their quasimomentum
(rapidity), $v_n(k)$ is their velocity, and $s_n(k)$ their contribution to the entanglement entropy.  The quasiparticle prediction  for
the entanglement entropy holds true in the space-time scaling limit, i.e. $t$,$\ell\to\infty$ with the ratio $t/\ell$ fixed. When a
maximum quasiparticle velocity $v_M$ exists (e.g., as a consequence of the Lieb-Robinson bound), Eq.~\ref{qp} predicts
that for $t \leq \ell/(2v_M)$, $S$ grows linearly in time. Conversely, for $t> \ell/(2v_M)$, only the second term survives and
the entanglement is extensive in the subsystem size, i.e., $S\propto\ell$.  The validity of Eq.~\ref{qp}
has been  tested both analytically and numerically in free-fermion and free-boson models  
~\cite{calabrese-2005,fagotti-2008,ep-08,nr-14,kormos-2014,leda-2014,collura-2014,bhy-17,hbmr-17,buyskikh-2016,fnr-17,mm-20,knn-19,dat-19} and in many interacting 
integrable models~\cite{alba-2016,PVCP18_I, PVCP18_II,MBPC17,modak.2019,dmcf-06}.

Hence, in Fig.~\ref{fig4}(b) we also show the quasi-particle  results using solid lines for $\Delta_f=0$ and for different initial product state. Also since, our tDMRG results are for open boundary condition, we replace $t\to t/2$ in Eq.~\eqref{qp}.  
It seems that entanglement growth rate obtained within this semi-classical picture matches reasonably well with our numerical results even though our simulation is for finite size system and $\Delta_f=0.2$.  
Note that even though the correlations between the entanglement growth \emph{rate} and the positioning of the non-analytic point $t^{*}$ is extremely apparent when we quench from the localized phase to the delocalized phase, but entanglement dynamics is featureless when quench is performed from the highly entangled delocalized phase to the localized phase.  However,  signature of the DQPT remains present in  the temporal variation of $f(t)$ [see Fig.~\ref{fig1}].

\section{Effects of interaction}
For all our previous calculations, we set the interaction term in the Hamiltonian $\hat{H}$ (see Eq.~\eqref{nonint_model}) to zero, i.e. $V=0$. Here we investigate the effect of interactions. While we know the Hamiltonian in ~\eqref{nonint_model} exhibits localization-delocalization transition at $\Delta=2$ in the non-interacting limit, the effect of interaction usually tends to delocalized the system. However, it has been shown that even in the presence of interaction for the sufficiently large value of the disorder strength (in this case the strength of incommensurate potential) localization-delocalization transition takes place, which is also known as ergodic to Many-body localization transition ~\cite{Iyer13}. 

\begin{figure}
\includegraphics[width=0.48\textwidth]{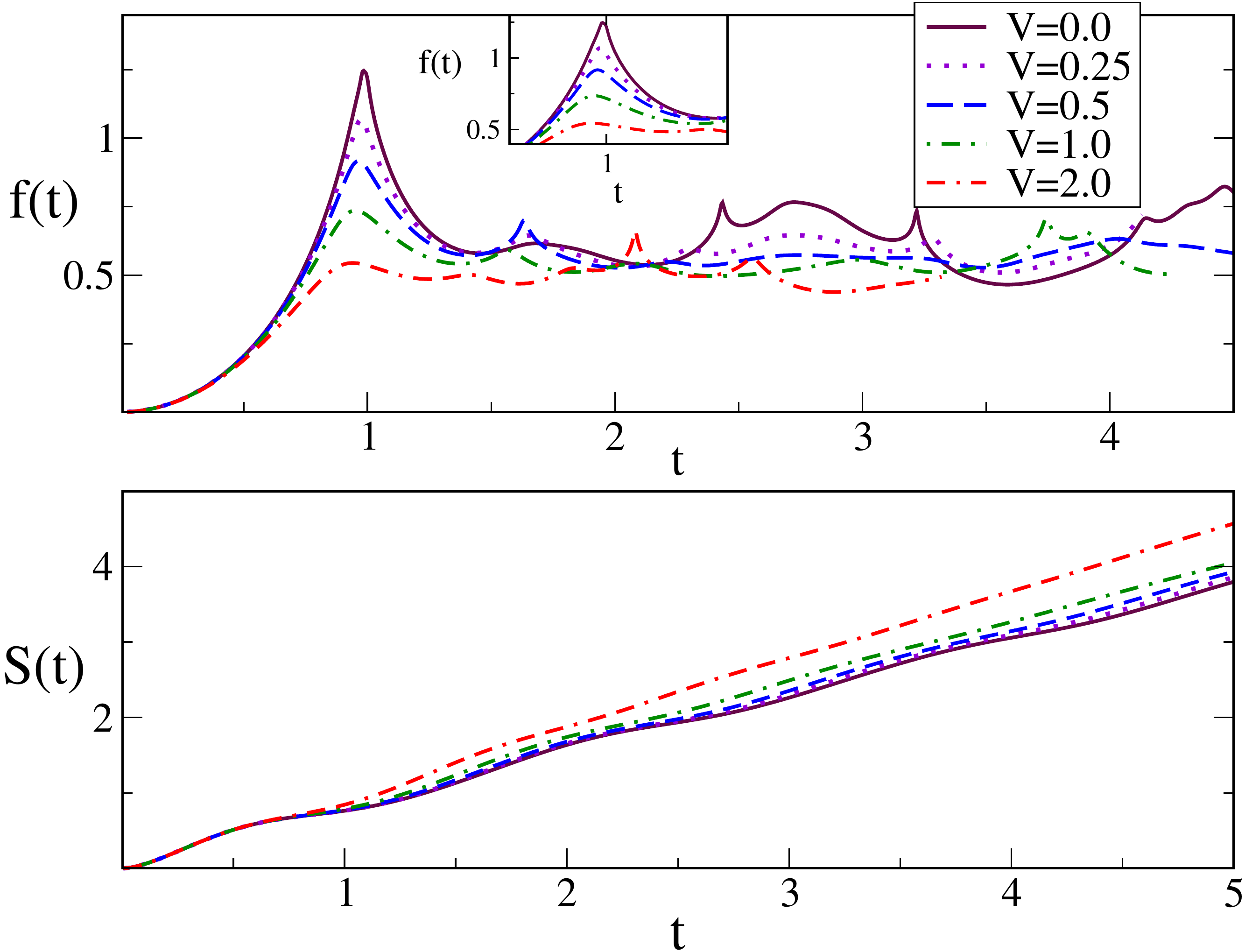}
\caption{\emph{Effects of interaction.} (a) Dynamical free energy, $f(t)$, variation with time, $t$,  for different interaction strengths, $V$, for $\Delta_i = 100$, and $\Delta_f = 0.2$. (b) shows entanglement dynamics for the same set of interaction strengths and system parameters.}
\label{fig5}
\end{figure}

Here, we perform a quench from the deep in to the many-body localized phase to the delocalized ergodic  phase  phase corresponding to $\Delta_i=100$ $\to $ $\Delta_f=0.2$ and $V\neq0$. 
We find that as we increase the interaction strength $V$, the signature  of the  non-analyticity in $f(t)$ slowly fades way as shown in fig.~\ref{fig5} (a). However interestingly  we also observe that the value of  $t^{*}$ becomes smaller with the increase of $V$. These results can also be complimented by the entanglement dynamics results in Fig.~\ref{fig5} (b), which show that indeed the entanglement growth is comparatively faster if we increase the interaction strength.

\section{Summary and Discussions}
The primary goal of this is work to understand the many-body aspects in the context of DQPT, for which we turned our focus to a less explored quasi-periodic system supporting a localization-delocalization transition. Our \emph{many-body} results indicate that DQPT is a generic feature of sudden quenches across the localization-delocalization transition point.

First we consider the non-interacting systems with finite particle densities, and investigate how the DQPT gets modified as a function the filling fraction when the system is quenched from the delocalized (localized) to the localized (delocalized) phase. Further investigation is followed in order to understand the effects of interaction. In the presence of interaction, single particle localized phase modifies to athermal MBL phase.  Here, we find a clear signature of non-analyticity in the dynamical free energy, as well. Since, unlike the  usual quantum phase transition (which involves only the ground states of the Hamiltonian),  the highly excited eigenstates of the system are also involved in MBL transition, we do not restrict ourselves to the only ground state of the pre-quench Hamiltonian -- We find that the signature of DQPT persists even for the highly excited states. 
 
While quenching from a low-entangled localized phase to high-entangled delocalized phase, we find that DQPT is accompanied with entanglement production. Crucially, following the trends of entanglement growth, we conclude that the positioning of the non-analytic points $t^{*}$ intrinsically depend on the \emph{rate} of entanglement growth. 

However,  the rate of entanglement growth does not solely determines the positions of $t^*$. Clearly, the onsets of $t^*$ do not require the quasi-particles to travel across the entire system and coming back to the initial point in the space. The time required for quasi-particles to do so is given by, $[t_{rev}\sim n L/(v_M)]$ with $n$ as positive integers ~\cite{modak.2020},  which is much larger (also scales linearly with system size $L$) than the transition times.

This is even more clear for the case when quenching is performed from the delocalized phase (initial state is highly entangled)  to the deep into localized phase, which is only associated with a negligible entanglement growth. However, even there the values of $t^{*}$ changes with the filling fraction. Hence, further investigations are required to understand the correlation between the entanglement growth and the DQPT. It will be interesting to explore the roles of local entanglement or global genuine multiparty entanglement. Even identifying the relations between quantum scrambling and DQPT is also another potential future direction~\cite{chen.2020}.

\section{Acknowledgements}
The authors acknowledges the support of IOE start-up grant.
RM acknowledges the support of DST-Inspire fellowship, by the Department of Science and Technology, Government of India. Super computing Facilities at IIT (BHU) were used to perform numerical computations.


\begin{thebibliography}{100}

\bibitem{Eisert15} J. Eisert, M. Friesdorf, and C. Gogolin, \emph{Quantum many-body systems out of equilibrium}, Nat. Phys. {\bf 11}, 124 (2015).

\bibitem{Heyl18} M. Heyl, \emph{Dynamical quantum phase transitions: a review}, Rep. Prog. Phys. {\bf 81}, 054001 (2018).

\bibitem{Heyl13} M. Heyl, A. Polkovnikov, and S. Kehrein, \emph{Dynamical quantum phase transitions in the transverse-field Ising model}, Phys. Rev. Lett. {\bf 110}, 135704 (2013).
\bibitem{Sedlmayr19} N. Sedlmayr, Acta Phys. Pol. A 135, 1191 (2019).
\bibitem{Karrasch13} C. Karrasch and D. Schuricht, \emph{Dynamical phase transitions after quenches in nonintegrable models}, Phys. Rev. B {\bf 87}, 195104 (2013).
\bibitem{Andraschko14} F. Andraschko and J. Sirker, \emph{Dynamical quantum phase transitions and the Loschmidt echo: A transfer matrix approach}, Phys. Rev. B {\bf 89}, 125120 (2014).
\bibitem{Halimeh18} J. C. Halimeh, M. Van Damme, V. Zauner-Stauber, and L.Vanderstraeten, \emph{Quasiparticle origin of dynamical quantum phase transitions}, Phys. Rev. Research {\bf 2}, 033111 (2020).
\bibitem{Mishra18} U. Mishra, R. Jafari, and A. Akbari, J. Phys. A: Math. Theor. {\bf 53}, 375301 (2020).
\bibitem{Shpielberg18} O. Shpielberg, T. Nemoto, and J. Caetano, \emph{Topological invariants and phase diagrams for one-dimensional two-band non-Hermitian systems without chiral symmetry}, Phys. Rev. E {\bf 98}, 052116 (2018).
\bibitem{Srivastav19} V. Srivastav, U. Bhattacharya, and A. Dutta, \emph{Dynamical quantum phase transitions in extended toric-code models}, Phys. Rev. B {\bf 100}, 144203 (2019).
\bibitem{Zunkovic18} B. Zunkovic, M. Heyl, M. Knap, and A. Silva, \emph{Dynamical quantum phase transitions in spin chains with long-range interactions: Merging different concepts of nonequilibrium criticality}, Phys. Rev. Lett. {\bf 120}, 130601 (2018).
\bibitem{Hagymasi19} I. Hagym{\'a}si, C. Hubig, {\"O}. Legeza, and U. Schollwöck, \emph{Dynamical topological quantum phase transitions in nonintegrable models}, Phys. Rev. Lett. {\bf 122}, 250601 (2019).
\bibitem{Huang19} Y.-P. Huang, D. Banerjee, and M. Heyl, \emph{Dynamical quantum phase transitions in U(1) quantum link models}, Phys. Rev. Lett. {\bf 122}, 250401 (2019).

\bibitem{Soriente19} M. Soriente, R. Chitra, and O. Zilberberg, \emph{Distinguishing phases using the dynamical response of driven-dissipative light-matter systems}, Phys. Rev. A {\bf 101}, 023823 (2020).
\bibitem{Cao19} K. Cao, W. Li, M. Zhong, and P. Tong, \emph{Influences of weak disorder on dynamical quantum phase transitions of anisotropic XY chain}, Phys. Rev. B {\bf 102}, 014207 (2020).
\bibitem{Gurarie19} V. Gurarie, \emph{Dynamical quantum phase transitions in the random field Ising model}, Phys. Rev. A {\bf 100}, 031601(R) (2019).
\bibitem{Abdi19} M. Abdi, \emph{Dynamical quantum phase transition in Bose-Einstein condensates}, Phys. Rev. B {\bf 100}, 184310 (2019).
\bibitem{Haldar20} S. Haldar, S. Roy, T. Chanda, A. Sen(De), and U. Sen, \emph{Multipartite entanglement at dynamical quantum phase transitions with nonuniformly spaced criticalities}, Phys. Rev. B {\bf 101}, 224304 (2020).

\bibitem{Vajna15} S. Vajna and B. D{\'o}ra, \emph{Topological classification of dynamical phase transitions}, Phys. Rev. B {\bf 91}, 155127 (2015).
\bibitem{Schmitt15} M. Schmitt and S. Kehrein, \emph{Dynamical quantum phase transitions in the Kitaev honeycomb model}, Phys. Rev. B {\bf 92}, 075114 (2015).

\bibitem{Jafari17} R. Jafari and H. Johannesson, \emph{Loschmidt echo revivals: Critical and noncritical}, Phys. Rev. Lett. {\bf 118}, 015701 (2017).
\bibitem{Sedlmayr18} N. Sedlmayr, P. J{\"a}ger, M. Maiti, and J. Sirker, \emph{Bulk-boundary correspondence for dynamical phase transitions in one-dimensional topological insulators and superconductors}, Phys. Rev. B {\bf 97}, 064304 (2018).
\bibitem{Jafari18} R. Jafari, H. Johannesson, A. Langari, and M. A. Martin-Delgado, \emph{Quench dynamics and zero-energy modes: The case of the Creutz model}, Phys. Rev. B {\bf 99}, 054302 (2018).
\bibitem{Zache19} T. V. Zache, N. Mueller, J. T. Schneider, F. Jendrzejewski, J. Berges, and P. Hauke, \emph{Dynamical topological transitions in the massive schwinger model with a ~$\theta$~ term}, Phys. Rev. Lett. {\bf 122}, 050403 (2019).


\bibitem{Sharma14} S. Sharma, A. Russomanno, G. E. Santoro, and A. Dutta, \emph{Loschmidt echo and dynamical fidelity in periodically driven quantum systems}, Europhys. Lett. {\bf 106}, 67003 (2014).
\bibitem{Yang19} K. Yang, L. Zhou, W. Ma, X. Kong, P. Wang, X. Qin, X. Rong, Y. Wang, F. Shi, J. Gong, and J. Du, \emph{Floquet dynamical quantum phase transitions}, Phys. Rev. B {\bf 100}, 085308 (2019).

\bibitem{Abeling16} N. O. Abeling and S. Kehrein, \emph{Quantum quench dynamics in the transverse field Ising model at nonzero temperatures}, Phys. Rev. B {\bf 93}, 104302 (2016).
\bibitem{Mera17} B. Mera, C. Vlachou, N. Paunkovi´c, V. R. Vieira, and O.Viyuela, \emph{Dynamical phase transitions at finite temperature from fidelity and interferometric Loschmidt echo induced metrics}, Phys. Rev. B {\bf 97}, 094110 (2017).
\bibitem{Bhattacharya17} U. Bhattacharya, S. Bandyopadhyay, and A. Dutta, \emph{Mixed state dynamical quantum phase transitions}, Phys. Rev. B {\bf 96}, 180303(R) (2017).
\bibitem{Heyl17} M. Heyl and J. C. Budich, \emph{Dynamical topological quantum phase transitions for mixed states}, Phys. Rev. B {\bf 96}, 180304(R) (2017).
\bibitem{Lang18a} J. Lang, B. Frank, and J. C. Halimeh, \emph{Dynamical quantum phase transitions: A geometric picture}, Phys. Rev. Lett. 121, 130603 (2018).
\bibitem{Lang18b} J. Lang, B. Frank, and J. C. Halimeh, \emph{Concurrence of dynamical phase transitions at finite temperature in the fully connected transverse-field Ising model}, Phys. Rev. B {\bf 97}, 174401 (2018).
\bibitem{Kyaw18} T. H. Kyaw, V. M. Bastidas, J. Tangpanitanon, G. Romero, and L.-C. Kwek, \emph{Dynamical quantum phase transitions and non-Markovian dynamics}, Phys. Rev. A {\bf 101}, 012111 (2020).

\bibitem{Lewenstein07} M Lewenstein, A Sanpera, V Ahufinger, B Damski, A Sen, and U Sen, \emph{Ultracold atomic gases in optical lattices: mimicking condensed matter physics and beyond}, Adv. Phys. {\bf 56}, 243 (2007).
\bibitem{Lewenstein12} M. Lewenstein, A. Sanpera, and V Ahufinger, \emph{Ultracold Atoms in Optical Lattices: Simulating quantum many-body systems}, Oxford University Press (2012).
\bibitem{Georgescu14} I.M. Georgescu, S. Ashhab, and Franco Nori, \emph{Quantum simulation}, Rev. Mod. Phys. {\bf 86}, 153 (2014).

\bibitem{Smale19} S. Smale, P. He, B. A. Olsen, K. G. Jackson, H. Sharum, S. Trotzky, J. Marino, A. M. Rey, and J. H. Thywissen, \emph{Observation of a transition between dynamical phases in a quantum degenerate Fermi gas}, Sci. Adv. {\bf 5}, eaax1568 (2019).
\bibitem{Jurcevic17} P. Jurcevic, H. Shen, P. Hauke, C. Maier, T. Brydges, C. Hempel, B. P. Lanyon, M. Heyl, R. Blatt, and C. F. Roos, \emph{Direct observation of dynamical quantum phase transitions in an interacting many-body system}, Phys. Rev. Lett. 119, 080501 (2017).

\bibitem{Anderson58} P. W. Anderson, \emph{Absence of Diffusion in Certain Random Lattices}, Phys. Rev. {\bf 109}, 1492 (1958).
\bibitem{Abrahams79} E. Abrahams, P. W. Anderson, D. C. Licciardello, and T. V. Ramakrishnan, \emph{Scaling theory of localization: Absence of quantum diffusion in two dimensions}, Phys. Rev. Lett. {\bf 42}, 673 (1979).

\bibitem{Abani17} D. A. Abanin and Z. Papic,´ \emph{Recent progress in many-body localization}, Ann. Phys. (Berlin) {\bf 529}, 1700169 (2017)
\bibitem{Alet18} F. Alet and N. Laflorencie, \emph{Many-body localization: An introduction and selected topics}, C. R. Phys.c {\bf 19}, 498 (2018).
\bibitem{Abanin19} D. A. Abanin, E. Altman, I. Bloch, and M. Serbyn, \emph{Colloquium: Many-body localization, thermalization, and entanglement}, Rev. Mod. Phys. {\bf 91}, 021001 (2019).
\bibitem{Nandkishore6} R. Nandkishore and D. A. Huse, \emph{Many body localization and thermalization in quantum statistical mechanics}, Annual Review of Condensed Matter Physics, {\bf 6}, 15 (2015).


\bibitem{Schreiber15} M. Schreiber, S. S. Hodgman, P. Bordia, H. P. L{\"u}schen, M. H. Fischer, R. Vosk, E. Altman, U. Schneider, and I. Bloch, \emph{Observation of many-body localization of interacting fermions in a quasirandom optical lattice}, Science {\bf 349}, 842 (2015).
\bibitem{Luschen17} H. P. L{\"u}schen, P. Bordia, S. Scherg, F. Alet, E. Altman, U. Schneider, and I. Bloch, \emph{Observation of slow dynamics near the many-body localization transition in one-dimensional quasiperiodic systems}, Phys. Rev. Lett. {\bf 119}, 260401 (2017).
\bibitem{Luschen18} H. P. L{\"u}schen, S. Scherg, T.
Kohlert, M. Schreiber, P. Bordia, X. Li, S. Das Sarma, and I. Bloch, \emph{Single-particle mobility edge in a one-dimensional quasiperiodic optical lattice}, Phys. Rev. Lett. {\bf 120}, 160404 (2018).

\bibitem{Yang17} C. Yang, Y. Wang, P. Wang, X. Gao, and S. Chen, \emph{Dynamical signature of localization-delocalization transition in a one-dimensional incommensurate lattice}, Phys. Rev. B {\bf 95}, 184201 (2017).

\bibitem{Aubry80} S. Aubry and G. Andr{\'e}, \emph{Analyticity breaking and Anderson localization in incommensurate lattices}, Ann. Israel Phys. Soc {\bf 3}, 133 (1980).
\bibitem{Modugno09} M. Modugno, \emph{Exponential localization in one-dimensional quasi-periodic optical lattices}, New J. Phys. {\bf 11}, 033023 (2009).

\bibitem{Machida86} K. Machida and M. Fujita, \emph{Quantum energy spectra and one-dimensional quasiperiodic systems}, Phys. Rev. B {\bf 34}, 7367 (1986).
\bibitem{Geisel91} T. Geisel, R. Ketzmerick, and G. Petschel, \emph{New class of level statistics in quantum systems with unbounded diffusion}, Phys. Rev. Lett. {\bf 66}, 1651 (1991).
\bibitem{Roscilde08} T. Roscilde, \emph{Bosons in one-dimensional incommensurate superlattices}, Phys. Rev. A {\bf 77}, 063605 (2008).
\bibitem{Roux08} G. Roux, T. Barthel, I. P. McCulloch, C. Kollath, U. Schollw{\"o}ck, and T. Giamarchi, \emph{Quasiperiodic Bose-Hubbard model and localization in one-dimensional cold atomic gases}, Phys. Rev. A {\bf 78}, 023628 (2008).
\bibitem{Deng08} X. Deng, R. Citro, A. Minguzzi, and E. Orignac, \emph{Phase diagram and momentum distribution of an interacting Bose gas in a bichromatic lattice}, Phys. Rev. A {\bf 78}, 013625 (2008).
\bibitem{Albert10} M. Albert and P. Leboeuf, \emph{Localization by bichromatic potentials versus Anderson localization}, Phys. Rev. A {\bf 81}, 013614 (2010).
\bibitem{Cai10} X. Cai, S. Chen, and Y. Wang, \emph{Superfluid-to-Bose-glass transition of hard-core bosons in a one-dimensional incommensurate optical lattice}, Phys. Rev. A {\bf 81}, 023626 (2010).
\bibitem{He12} K. He, I. I. Satija, C. W. Clark, A. M. Rey, and M. Rigol, \emph{Noise correlation scalings: Revisiting the quantum phase transition in incommensurate lattices with hard-core bosons}, Phys. Rev. A {\bf 85}, 013617 (2012).
\bibitem{Gullo15} N. Lo Gullo and L. Dell'Anna, \emph{Spreading of correlations and Loschmidt echo after quantum quenches of a Bose gas in the Aubry-Andr{\'e} potential}, Phys. Rev. A {\bf 92}, 063619 (2015).

\bibitem{Iyer13} S. Iyer, V. Oganesyan, G. Refael, and
D. A. Huse, \emph{Many-body localization in a quasiperiodic system}, Phys. Rev. B {\bf 87}, 134202 (2013).

\bibitem{Eisert10} J. Eisert, M. Cramer, and M. B. Plenio, \emph{Colloquium: Area laws for the entanglement entropy}, Rev. Mod. Phys. {\bf 82}, 277 (2010).
\bibitem{Amic10} L. Amico, R. Fazio, A. Osterloh, and V. Vedral, \emph{Entanglement in many-body systems}, Rev. Mod. Phys. {\bf 80}, 517 (2008).

\bibitem{Islam15} R. Islam, R. Ma, P. M. Preiss, M. E. Tai, A. Lukin, M. Rispoli, and M. Greiner, \emph{Measuring entanglement entropy in a quantum many-body system}, Nature, {\bf 528}, 7580 (2015).
\bibitem{Kaufman16} A. M. Kaufman, M. E. Tai, A. Lukin, M. Rispoli, R. Schittko, P. M. Preiss, and M. Greiner, \emph{Quantum thermalization through entanglement in an isolated many-body system}, Science {\bf 353}, 794 (2016).




\bibitem{white-2004}
S.~R.~White and A.~E.~Feiguin, \emph{Real-Time evolution using the density matrix renormalization group}, Phys.\ Rev.\ Lett.\ {\bf 93}, 076401 (2004).

\bibitem{daley-2004}
A.~J.~Daley, C.~Kollath, U.~Schollock, and G.~Vidal, \emph{Time-dependent density-matrix renormalization-group using adaptive effective Hilbert spaces}, J.\ Stat.\ Mech.\ {\bf 2004}, 04005 (2004).

\bibitem{uli-2011}
U.~Schollw\"ock, 
\emph{The density-matrix renormalization group in the age of matrix product states}, Ann. Phys. {\bf 326}, 96 (2011).



\bibitem{calabrese-2005}
P.~Calabrese and J.~Cardy, \emph{Evolution of Entanglement Entropy in One-Dimensional Systems}, 
J. Stat. Mech. {\bf 2005},  04010 (2005).

\bibitem{alba-2016}
V.~Alba and P.~Calabrese, \emph{Entanglement and thermodynamics after a quantum quench in integrable systems}, Proc. Natl. Acad. Sci. {\bf 114}, 7947 (2017). 

\bibitem{fagotti-2008}
M.~Fagotti and P.~Calabrese,  
\emph{Evolution of entanglement entropy following a quantum quench: Analytic results for the XY chain in a transverse magnetic field}, Phys. Rev. A {\bf 78}, 010306 (2008). 

\bibitem{ep-08}
V. Eisler and I. Peschel, \emph{Entanglement in a periodic quench}, Ann. Phys. (Berlin) {\bf 17}, 410 (2008).

\bibitem{nr-14}
M. G.~Nezhadhaghighi and M. A. Rajabpour, \emph{Entanglement dynamics in short and long-range harmonic oscillators}, Phys. Rev. B {\bf 90}, 205438 (2014).

\bibitem{kormos-2014} M.~Kormos, L.~Bucciantini, and P.~Calabrese, \emph{Stationary entropies after a quench from excited states in the Ising chain}, Europhys. Lett. {\bf 107}, 40002 (2014). 


\bibitem{leda-2014} L. Bucciantini, M. Kormos, and P. Calabrese, \emph{Quantum quenches from excited states in the Ising chain}, J. Phys. A {\bf 47}, 175002 (2014).

\bibitem{collura-2014} M.~Collura, M.~Kormos, and P.~Calabrese, \emph{Stationary entropies following an interaction quench in $1D$ Bose gas}, J. Stat. Mech. {\bf 2014}, 01009 (2014).

\bibitem{bhy-17}
E. Bianchi, L. Hackl, and N. Yokomizo, Linear growth of the entanglement entropy and the Kolmogorov-Sinai rate,
JHEP {\bf 03}, 25 (2018).

\bibitem{hbmr-17}
L. Hackl, E. Bianchi, R. Modak, and M. Rigol, \emph{Entanglement production in bosonic systems: Linear and logarithmic growth}, Phys. Rev. A {\bf 97}, 032321 (2018).

\bibitem{buyskikh-2016}
A. S. Buyskikh, M. Fagotti, J. Schachenmayer, F. Essler, and A. J. Daley,
\emph{Entanglement growth and correlation spreading with variable-range interactions in spin and fermionic tunnelling models}, Phys.\ Rev.\ A {\bf 93}, 053620 (2016). 

\bibitem{fnr-17}
I. Frerot, P. Naldesi, and T. Roscilde,  \emph{Multi-speed prethermalization in spin models with power-law decaying interactions}, Phys. Rev. Lett. {\bf 120}, 050401 (2018).

\bibitem{mm-20}
M. R. M. Mozaffar  and A. Mollabashi, \emph{Entanglement evolution in Lifshitz-type scalar theories}, JHEP {\bf 01}, 137 (2019).

\bibitem{knn-19}
K.-Y. Kim, M. Nishida, M. Nozaki, M. Seo, Y. Sugimoto, and A. Tomiya, \emph{Entanglement after quantum quenches in Lifshitz scalar theories}, J. Stat. Mech. (2019) 093104.


\bibitem{dat-19}
G. Di Giulio, R. Arias, and E. Tonni, \emph{Entanglement Hamiltonians in 1D free lattice models after a global quantum quench}, J. Stat. Mech. {\bf 2019}, 123103 (2019).


\bibitem{PVCP18_I} 
L. Piroli, E. Vernier, P. Calabrese, and B. Pozsgay, \emph{Integrable quenches in nested spin chains I: the exact steady states}, J. Stat. Mech. {\bf 2019},  063103 (2019).

\bibitem{PVCP18_II} 
L. Piroli, E. Vernier, P. Calabrese, and B. Pozsgay, \emph{Integrable quenches in nested spin chains II: fusion of boundary transfer matrices},J. Stat. Mech. {\bf 2019}, 063104 (2019).

\bibitem{MBPC17} 
M. Mesty\'an, B. Bertini, L. Piroli, and P. Calabrese, \emph{Exact solution for the quench dynamics of a nested integrable system}, J. Stat. Mech. {\bf 2017},  083103 (2017).

\bibitem{modak.2019}
R. Modak, L. Piroli, and P. Calabrese, \emph{Correlation and entanglement spreading in nested spin chains}, J. Stat. Mech. {\bf 2019}, 093106 (2019).

\bibitem{dmcf-06}
G. De Chiara, S. Montangero, P. Calabrese, and R. Fazio,  \emph{Entanglement entropy dynamics in Heisenberg chains}, J. Stat. Mech. {\bf 2006}, 03001 (2006).
\bibitem{modak.2020}
R. Modak and V. Alba and P. Calabrese, \emph{Entanglement revivals as a probe of scrambling in finite quantum systems},
	J. Stat. Mech. {\bf 2020}, 083110 (2020). 
\bibitem{chen.2020}	B. Chen, X. Hou, F. Zhou, P. Qian,H. Shen, and N. Xu1, \emph{Detecting dynamical quantum phase transition via out-of-time-order correlations in a solid-state quantum simulator},   arXiv:2001.06333 (2020).

\end{thebibliography}
\end{document}